# Microwave performance of high-density bulk MgB$_2$

A. T. Findikoglu,[a] A. Serquis, L. Civale, X. Z. Liao, Y. T. Zhu, M. E. Hawley, and F. M. Mueller

Superconductivity Technology Center, MS T004
Materials Science and Technology Division
Los Alamos National Laboratory, Los Alamos, New Mexico 87545

V. F. Nesterenko, and Y. Gu

Department of Mechanical and Aerospace Engineering
University of California, San Diego
La Jolla, California 92093

## Abstract

We have performed microwave measurements on superconducting hot-isostatically-pressed (HIPed) bulk MgB$_2$ using a parallel-plate resonator technique. The high density and strength of the HIPed material allowed preparation of samples with mirror-like surfaces for microwave measurements. The microwave surface resistance decreased by about 40% at 20 K when the root-mean-square surface roughness was reduced from 220 nm to 110 nm through surface-polishing and ion-milling. The surface resistance was independent of surface microwave magnetic field at least up to 4 Oe and below 30 K. We attribute this behavior, and the overall low surface resistance (~0.8 m$\Omega$ at 10 GHz and 20 K), to the high density of our samples and the absence of weak links between grains.

---

[a] Electronic mail: findik@lanl.gov

The discovery of superconductivity at 39 K in $MgB_2$ has generated interest for its commercial use in electric power, microwave, and electronics applications.[1] For electric power applications, low-cost production potential of wires using easily-scalable techniques such as powder-in-tube, and high critical currents due to apparent absence of weak-link behavior at grain boundaries have been the main economic drivers.[2,3] Tunneling results show that the $MgB_2$ superconductor has a well-developed s-wave energy gap.[4] Also, superconductor-normal-superconductor planar junctions that were prepared using localized ion damage in thin-film $MgB_2$ exhibit conventional Josephson effect.[5] These results have raised hope for the use of such thin films of $MgB_2$ in electronics at 10-30 K, the temperature range achievable using closed-cycle cryocoolers.

For microwave applications, $MgB_2$ holds promise for filling a niche between low-transition-temperature (low-$T_c$) superconductors such as Nb and high-transition-temperature (high-$T_c$) oxides such as $YBa_2Cu_3O_{7-\delta}$ (YBCO). The fact that $MgB_2$ is a very good metal with relatively high electrical conductivity and the possibility to prepare samples with no weak-links represent significant advantages with respect to oxide high-$T_c$ superconductors. The results in the last two years on microwave properties of both bulk and thin-film $MgB_2$ show a wide spread of critical parameters such as the microwave surface resistance $R_s$, and its nonlinearities, indicating that some of the early results may not be indicative of intrinsic properties.[6,7,8,9] The most recent results on high-quality thin-films have been especially encouraging:[10] They demonstrate classic s-wave-like order parameter, which leads to an exponential drop of surface resistance below about $T_c/2$ in optimized samples, and very low residual surface resistance of 19 $\mu\Omega$ measured at 5 K and 7.2 GHz.

In this letter, we report on encouraging results on microwave properties of bulk $MgB_2$, prepared by a new DMCUP (dense material cooled under pressure) process based on hot isostatic pressing (HIP), the details of which have been published elsewhere.[11,12] The density of the as-prepared 30-mm-diameter $MgB_2$ sample was 2.56 g/cc. This value lies between the theoretical density of 2.625 g/cc based on x-ray measurements,[13] and the theoretical density of 2.55 g/cc reported in ref. [14]. The elastic moduli of this sample measured using resonant ultrasound spectroscopy method[15] are: 127.5 GPa for the bulk modulus, 245.2 GPa for the Young's modulus, and 0.18 for the Poisson ratio (averaged data for 3 different specimens taken from the same sample).[16] These values of density and elastic moduli are consistent with a low porosity in the sample of less than 3%.

This HIPing process for $MgB_2$ is promising on two counts: (i) It employs a practical HIP procedure with a pressure range (~200 MPa) accessible by existing techniques. Thus, it allows economical preparation of samples of large size and complex shape such as microwave cavities. (ii) It produces highly-dense, low-porosity, crack-free samples which allow high-precision machining.[15] These characteristics, combined with the inherent non-weak-link behavior of the material lead to overall enhanced properties; such as full magnetization in the superconducting state, high critical current density ($J_c$) of about 1 MA/cm$^2$ below 20 K, and as described below, improved microwave properties.

We have measured microwave properties of our samples using a parallel plate configuration.[17] The details of our parallel plate measurement technique have been published previously.[18] This technique allows direct measurement of the surface resistance, and also provides information on its power dependence.[19] Figure 1 shows microwave surface resistance vs. temperature for a pair of 10 mm x 8 mm x 1 mm plates of $MgB_2$ cut from the 30-mm-diameter sample that was prepared using the HIPing process. The same pair of plates were measured after: (i) coarse-polishing (dotted line) using dry silicon carbide 1200/4000 sand-paper, (ii) fine-polishing (dashed line) using diamond paste with 0.1-µm particle size, and (iii) finally ion-etching (continuous line) using 750 eV Ar$^+$ ions, aligned at 45$^o$ to sample surface normal, with a flux of 6x10$^{13}$/cm$^2$s for five minutes (which we estimate should remove ~4 nm of surface material). Our measurements were performed around 11 GHz. This frequency was determined by the sample size, and the $R_s$ values were scaled to 10 GHz assuming a quadratic dependence on frequency. For comparison, we have also included data on oxygen-free-high-conductivity copper (straight line), polycrystalline YBCO (short-long dashed line),[6] and biaxially-textured YBCO with out-of-plane c-axis rocking curve of less than $2^o$, and in-plane mosaic spread of $7^o$ (squares)[19]. Below 30 K, the surface resistance of our $MgB_2$ samples lies between polycrystalline YBCO and biaxially-textured YBCO.

We observe two important points regarding HIPed $MgB_2$. First, the well-connected, low-porosity material allows for surface polishing and corresponding reduction in $R_s$ with surface smoothness. After polishing, we measured root-mean-square surface roughness using atomic force microscopy. The surface roughness is 220 nm after coarse-polishing and 110 nm after fine-polishing and ion-milling. The fact that similar reduction in $R_s$ is observed at both high- and low-temperature extremes (5 and 30 K) supports the hypothesis that this reduction is dominated by the geometric effect of reduced surface area seen by the microwaves.[6] Instead, the removal of lossy extrinsic material would have led to much larger reduction at lower temperatures (i.e., drop in residual surface resistance), whereas effects dominated by removal of degraded superconducting material with reduced transition temperature would have caused larger reduction at higher temperatures. In contrast to a previous report,[20] we did not observe any significant change in $R_s$ with ion-etching, except for slight reduction due to some additional surface area reduction. We do not expect further significant reduction in $R_s$ through surface smoothening because the microwave fields will sense the surface of a superconductor only within a penetration depth, and the surface roughness of our fine-polished and ion-milled material is 110 nm, of the order of the magnetic penetration depth in $MgB_2$.

Secondly, as shown in Fig. 1, we did not observe any power dependence of $R_s$ at 5, 20 and 30 K in the microwave magnetic field ($H_{rf}$) range of 0.2 Oe (small filled circles) to 4 Oe (large unfilled circles). The power-handling capacity is a critical parameter for many applications of superconductors.[19,21] The initial enthusiasm regarding large-scale and inexpensive microwave applications of YBCO material that operate at liquid-nitrogen temperatures has somewhat cooled over the years mainly because of the observation of unconventional superconductivity that causes slower-than-exponential drop of $R_s$ below $T_c/2$ and the detrimental effects of weak-links present in not only polycrystalline and biaxially-textured, but also epitaxial thin films.[22] The presence of weak-links and the strong dependence of the associated Josephson current on the misalignment angle of the grains lead to flux penetration into polycrystalline and biaxially-textured samples at very low current levels.[23] This in turn leads to high and nonlinear microwave losses at very low dc and microwave magnetic field levels (<1 Oe).[19] For example, high-quality epitaxial YBCO films on sapphire show low surface resistance (less than 100 μΩ at 8.1 GHz below 70 K) and onset of nonlinearity at a fairly high microwave magnetic field level of 10 to 100 Oe.[24] But, that field level is still about two orders of magnitude smaller than the thermodynamic critical field $H_C$. In contrast, as shown in Fig. 1, even some of the best-quality biaxially-oriented YBCO films have shown significant increases in $R_s$ in the 0.2 - 4 Oe range (small filled rectangles and large unfilled rectangles) at 4 K and 77 K.[19] It is important to note that these biaxially textured YBCO films can support dc critical current densities approaching that of epitaxial films (~1 MA/cm$^2$ at 77 K) corresponding to surface magnetic fields in the range 10-100 Oe. However, the small-percentage weak-link regions of the material surface with low critical current density, which are percolatively shunted at low frequencies, are sampled by the microwave fields. In other words, power-dependent microwave measurements are sensitive probes of weak links in superconductors.

To investigate the magnetic properties and the weak-link behavior of our bulk samples further, we performed magnetization measurements on a 0.5mm x 1 mm x 4 mm specimen cut from one of the plates. Using a SQUID magnetometer, we measured the magnetization (M) as a function temperature (T) and magnetic field (H) in zero-field-cooling (ZFC) and field-cooling (FC) conditions. The ZFC values of susceptibility $4\pi M/H$ at T = 5 K in the H range of 1 - 1000 Oe are shown in Fig. 2. Within experimental resolution, the susceptibility is -1 up to about 200 Oe, indicating that the screening currents flow through the whole sample before one reaches the lower critical field $H_{C1}$ (the hatched region labeled $H_{C1}$ covers the range of values reported in the literature for $MgB_2$). In the left inset are the M(T) data, where both the ZFC and the FC curves are shown. The onset of transition temperature $T_c$ (38.5 K) is slightly lower than other bulk samples reported in the literature, suggesting some strain in the sample.[14] The ZFC curves show a sharp transition ($\Delta T$ [10-90%] is less than 1 K) and there is no indication of weak-link behavior within our experimental resolution up to about $H_{C1}$. FC values close to zero are indicative of a large pinning in this sample. The right inset of Fig. 2 displays the irreversibility field ($H_{irr}$) of the same sample as a function of temperature as determined by magnetization, consistent with the recently published data in the literature.[25]

The work done on conventional superconductors such as Nb and $Nb_3Sn$ have shown that carefully prepared radio-frequency (rf) cavities could sustain fields much above $H_{C1}$, up to and in some instances above the thermodynamic critical field $H_C$.[23] Also, through improvements in material purity and surface preparation, the residual rf surface resistance values have been brought down below a few $n\Omega$.[26] Such optimized samples unequivocally show BCS-theory-predicted exponential drop of $R_s$ below about $T_c/2$. The promise the $MgB_2$ material holds for rf and microwave applications stems from the fact that it appears to be more like conventional low-temperature superconductors, and less like high-temperature superconductors.[25,27] The results we presented above are a clear demonstration of lack of weak-links, and consequently of the potential for rf/microwave operation, in bulk samples prepared using a commercially viable process. Since these samples were optimized for low-frequency applications that require strong pinning, they are by design not "clean". Thus, we expect significant reduction in low temperature $R_s$ (i.e., residual $R_s$) once the material preparation process is optimized to avoid extrinsic materials and lossy pinning sites.

**Figure Captions**

Figure 1   Microwave surface resistance $R_s$ at 10 GHz vs temperature of: oxygen-free-high-conductivity Cu (straight line); polycrystalline [ref. 6] (long-short dashed line) and biaxially-textured (rectangles) $YBa_2Cu_3O_{7-\delta}$ [ref. 19]; coarse-polished (dotted line), fine-polished (dashed line), fine-polished and ion-milled (circles) bulk HIPed $MgB_2$, measured at microwave magnetic field level $H_{rf}$ of 0.2 Oe (small filled circles and rectangles), and 4 Oe (large unfilled circles and rectangles).

Figure 2   Magnetic susceptibility $4\pi M/H$ at 5 K of the fine-polished and ion-milled $MgB_2$ sample as a function of applied magnetic field H. The left inset shows $4\pi M/H$ as a function of temperature at H = 2, 20, 100 and 1000 Oe. The right inset shows the temperature dependence of the irreversibility field $H_{irr}$ of the same sample.

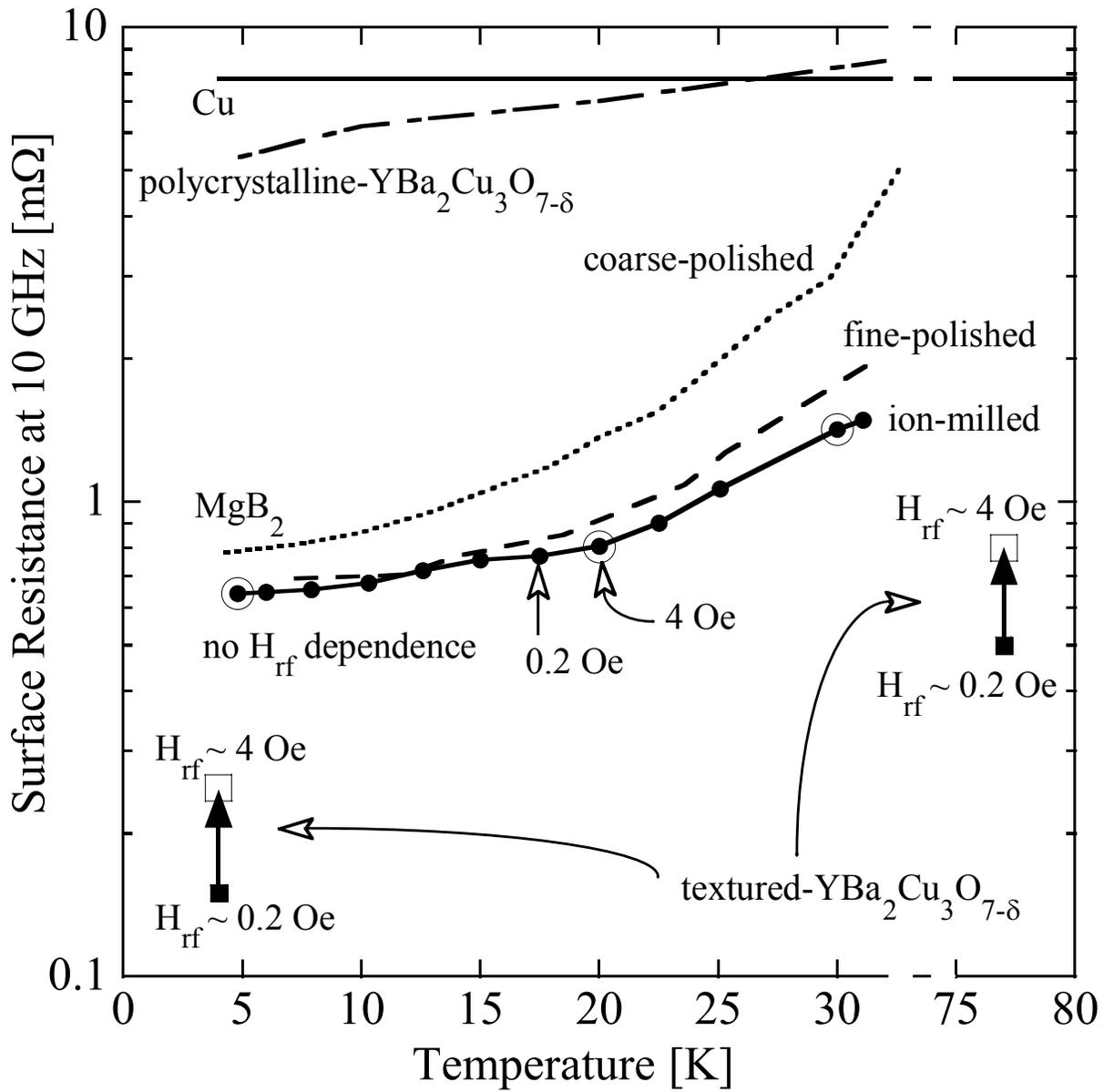

Figure 1  Findikoglu, et. al.

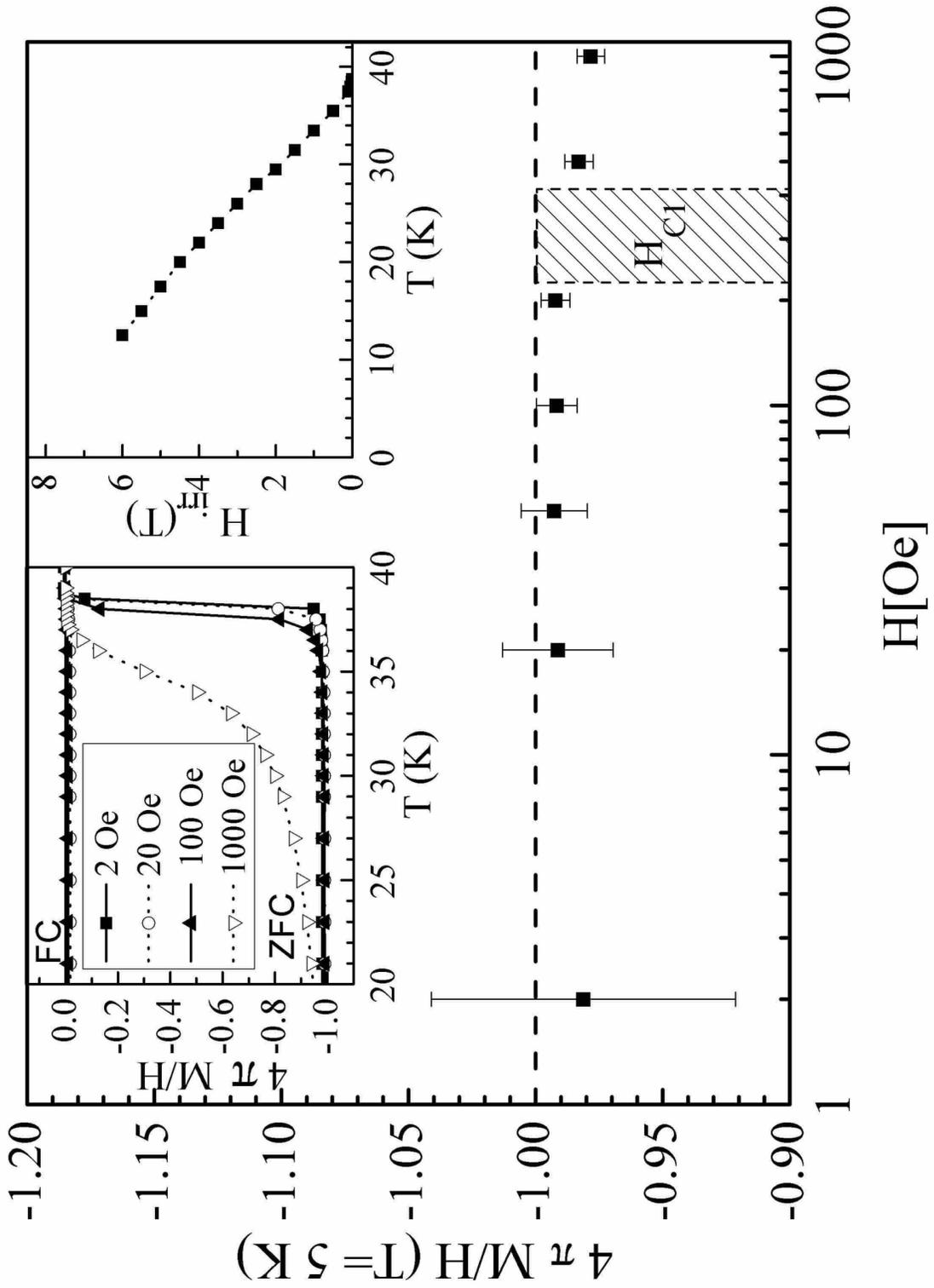

Figure 2                                           Findikoglu, et. al.